%% file: ms.tex
\documentclass[sigconf]{acmart}
\usepackage{xcolor}
\usepackage{booktabs}
\usepackage{multirow}
\usepackage{subcaption}
\usepackage{listings}
\usepackage{graphicx}
\usepackage[framemethod=tikz]{mdframed}
\mdfdefinestyle{mpdframe}{
    frametitlebackgroundcolor = black!15,
    frametitlerule            = true,
    roundcorner               = 5pt,
}
\lstdefinelanguage{diff}{
  numbers=left,
  numbersep=5pt,
  belowcaptionskip=1\baselineskip,
  breaklines=true,
  xleftmargin=\parindent,
  showstringspaces=false,
  basicstyle=\footnotesize\ttfamily,
  keywordstyle={\bfseries\color{green!40!black}},
  commentstyle={\itshape\color{purple!40!black}},
  stringstyle=\color{orange},
  morecomment=[f][\color{white!40!black}]{---\ },
  morecomment=[f][\color{white!40!black}]{+++\ },
  morecomment=[f][\color{white!40!black}]{@@},
  morecomment=[f][\color{green!50!black}]{+\ },
  morecomment=[f][\color{red!80!black}]{-\ },
}

\newcommand{\tool}{\textit{ActiveClean}}

\newcommand{\mycomment}[1]{}
\setcopyright{none}
\renewcommand\footnotetextcopyrightpermission[1]{}

\acmConference[ArXiv]{ArXiv}{December 2023}{Ames, USA}

\begin{document}

\title{ActiveClean: Generating Line-Level Vulnerability Data via~Active~Learning}



\author{Ashwin Kallingal Joshy}
\orcid{0000-0001-8236-5027}
\affiliation{%
  \institution{Iowa State University}
  \city{Ames}
  \state{IA}
  \country{USA}}
\email{ashwinkj@iastate.edu}
\author{Mirza Sanjida Alam}
\affiliation{%
  \institution{Iowa State University}
  \city{Ames}
  \state{IA}
  \country{USA}}
\email{sanjida@iastate.edu}
\author{Shaila Sharmin}
\affiliation{%
  \institution{Iowa State University}
  \city{Ames}
  \state{IA}
  \country{USA}}
\email{ssharmin@iastate.edu}
\author{Qi Li}
\affiliation{%
  \institution{Iowa State University}
  \city{Ames}
  \state{IA}
  \country{USA}}
\email{qli@iastate.edu}
\author{Wei Le}
\affiliation{%
  \institution{Iowa State University}
  \city{Ames}
  \state{IA}
  \country{USA}}
\email{weile@iastate.edu}

\input{abstract}

\maketitle

\input{intro}

\input{motivation}

\input{overview}

\input{features}

\input{learning}

\input{results}

\input{examples}

\input{threats}

\input{related}

\input{conclusion}

\bibliographystyle{ACM-Reference-Format}
\bibliography{Detangle_Fixes}
\end{document}

%% file: abstract.tex
\begin{abstract}
  Deep learning vulnerability detection tools are increasing in popularity and have been shown to be effective. These tools rely on large volume of high quality training data, which are very hard to get. Most of the currently available datasets provide \textit{function-level labels}, reporting whether a function is vulnerable or not vulnerable. However, for a vulnerability detection to be useful, we need to also know the lines that are relevant to the vulnerability. This paper makes efforts towards developing systematic tools and proposes \tool{} to generate the large volume of \textit{line-level} vulnerability data from commits. That is, in addition to function-level labels, it also reports which lines in the function are likely responsible for vulnerability detection. In the past, static analysis has been applied to clean commits to generate line-level data. Our approach based on \textit{active learning}, which is easy to use and scalable, provide a complementary approach to static analysis. We designed semantic and syntactic properties from commit lines and use them to train the model. We evaluated our approach on both Java and C datasets processing more than 4.3K commits and 119K commit lines. \tool{} achieved an F1 score between 70--74. Further, we also show that active learning is effective by using just 400 training data to reach F1 score of 70.23. Using \tool{}, we generate the line-level labels for the entire \texttt{FFMpeg} project in the Devign dataset, including 5K functions, and also detected incorrect function-level labels. We demonstrated that using our cleaned data, LineVul, a SOTA line-level vulnerability detection tool, detected 70 more vulnerable lines and 18 more vulnerable functions, and improved Top 10 accuracy from 66\% to 73\%.


\end{abstract}

%% file: intro.tex
\section{Introduction}\label{sec:CleanIntro}

Deep learning based vulnerability detection has shown effective~\cite{russell_automated_2018, li_vuldeepecker_2018,sestili_towards_2018,suneja_learning_2020} and is becoming important for mitigating the increasing~\cite{noauthor_zero_2023} number of vulnerabilities being exploited in the wild. Despite the rapid development of vulnerability detection models, obtaining high quality data has always remained a challenge~\cite{lin_software_2020}. The current vulnerability datasets contain false labels~\cite{herbold_fine-grained_2021, yang_is_2021}. Most of these datasets use \textit{function-level} labels and report whether a function is vulnerable~\cite{just_defects4j_2014,zhou_devign_2019}. Ideally, to be more useful, the tools should report \textit{line-level} vulnerabilities, e.g., which lines are responsible to produce the vulnerability. Lacking high quality line-level vulnerability datasets, building line-level vulnerability detection models can be significantly challenging. Later in the evaluation section of this paper, we showed how a line-level model can improve its performance after being trained with improved dataset.
In the past, most of the vulnerability detection datasets such as Devign~\cite{zhou_devign_2019}, MSR/Big-Vul~\cite{fan_cc_2020} and D2A~\cite{zheng_d2a_2021} are harvested from software patches in the code repositories~\cite{dallmeier_extraction_2007,le_goues_manybugs_2015,bohme_where_2017,jiang_bugbuilder_2023}. For example, Devign used keywords in the commit messages and then applied manual inspection to produce function-level labels. While we need a large number of examples for training vulnerability detection models, manually cleaning these patches to produce line-level labels is very expensive. In a recent effort, it took 45 authors and more than 6 months to generate line-level labels for 28 Java projects and 3,546 commits~\cite{herbold_fine-grained_2021}. D2A used the difference in warnings generated by static analysis tools from before and after the commit to produce \textit{line-level} datasets. While promising, this approach relies on static analysis tool's ability to accurately detect the vulnerability.

In this work, we make another attempt towards automatically generating line-level vulnerability data. We developed a tool called \tool{} that uses \textit{active learning} to train a machine learning model to clean up software patches. We selected the \textit{active learning} method because manually labeling the commit lines is expensive and active learning can help select the best and the minimum number of data points to label. We used machine learning based approach, because the trained model is easy to use and scalable to process a large number of data required by vulnerability detection. \tool{} takes input programs with vulnerable commits, and report whether a line is related to the vulnerability. \tool{} also can take as input existing function-level vulnerability datasets and return line-level labels. In that case, \tool{} also can report that a function-level label is incorrect when it finds that all the lines in the functions are labeled as \textit{vulnerability irrelevant}.




To train the model in \tool{}, we built a tool to automatically extract syntactic and semantic features from the commit lines. The features we design not only show characteristics of the commit lines themselves, but also how they interact with other commit lines as well as their surrounding code. Then we use \textit{query-by-committee} based active learning framework to first train a set of \textit{committee models} from a small set of existing line-level vulnerability dataset. During the second phrase of active learning, we query the \textit{committee} to select the best data to label to continue training the model on the new input data. 

In our evaluation, we used both Java and C datasets and processed a total of 4375 commits with 119K commit lines. Our results indicate that \tool{} is able to clean the commits and reduce vulnerability-irrelevant lines across all the benchmarks, with an F1 score between 70--74 for datasets with ground truth available. \tool{} also consistently outperformed the baselines models and different settings in ablation study. We also showed that active learning indeed brings efficiency. It took 400 additional training data to achieve an F1 score of 70.23, while baseline models need 2K training data. It took 400 additional training data to achieve an F1 score of 70.23, while baseline models only reached 67 even after 2K additional training data.

Using \tool{} we generated the line-level vulnerability data for \texttt{FFmpeg} in the Devign dataset. Using the help of \tool{}, we also detected incorrect function-level labels in the existing Devign dataset. We randomly sampled 50 false labels out of 468 reported by \tool{}, and we found 29 are indeed non-vulnerable, and 8 was not able to be confirmed manually. Using the cleaned dataset we produced, the SOTA vulnerability detection model LineVul detected 18 more vulnerable functions and 70 more vulnerable lines. It achieved 87 F1 score for function-level detection and 73\% Top 10 accuracy for line-level detection, compared to 83 F1 score and 66\% Top 10 accuracy achieved using the original dataset.

In summary, this paper produced a tool that is automatic and scalable to generate line-level vulnerability data. Although we have not achieved 100\% accuracy and recall, we demonstrated that using the data we cleaned, the vulnerability detection models can improve the performance. We hope that by providing the dataset such as line-level labels for FFmpeg, we can promote better tools in the area of deep learning for vulnerability detection. Meanwhile, we also see that many other applications such as automatic repair and fault localizations can benefit from the cleaned patches. In the future, we will continuously apply \tool{} to generate more such datasets.

Our contributions of the paper include:
\begin{enumerate}
  \item An automatic technique and scalable tool that can clean the patches and generate line-level vulnerability data,
  \item FFmpeg vulnerability dataset with line-level labels, consisting of 5K vulnerable functions with 9K labels,
  \item Feature engineering and an application of active learning as a novel method for this problem,
  \item A systematic evaluation to show our approach can improve the quality of the dataset and improve deep learning models for vulnerability detection.
\end{enumerate}

%% file: motivation.tex
\section{A Motivating Example}\label{sec:CleanMotivation}
\begin{figure}[ht]
  \centering
\begin{lstlisting}[language=diff]
@@ -239,9 +239,5 @@
  public double getRMS() {
-   double criterion =0;
-   for (int i = 0; i < rows; ++i) {
-     final double residual = residuals [i];
-     criterion += residual * residual / residualsWeights[i];
-   }
-   return Math.sqrt(criterion / rows);
+   return Math.sqrt(getChiSquare() / rows);
  }
@@ -254,11 +254,11 @@
  public double getChiSquare() {
    double chiSquare = 0;
-   double sqHalf = 0;
+   double chiSqHalf = 0;
    for (int i = 0; i < rows; ++i) {
      final double residual = residuals[i];
+     if residualsWeights[i] != 0
        chiSquare += residual * residual / residualsWeights[i];
    }
-   sqHalf = chiSquare / 2;
+   chiSqHalf = chiSquare / 2;
    return chiSquare;
  }
\end{lstlisting}
  \caption{An Example Adapted from Math 65 in \texttt{Defect4J} }\label{fig:CleanMath65}
\end{figure}

Defect4J~\cite{just_defects4j_2014} is a widely used benchmark where patches are manually minimized to contain the smallest changes for fixing bugs. However, due to the challenges of cleaning commits, the patches in Defect4J sometimes still contained changes that are irrelevant to the bugs~\cite{yang_is_2021}. In Figure~\ref{fig:CleanMath65}, we shown an example of uncleaned patch adapted from \textit{Math 65} in Defect4J.


Here, the patch affects two functions: \textit{getRMS} at line~2 and \textit{getChiSquare} at line~12. The commit consists of 12 lines (lines~3--9, 14, 15, 18, 21, and 22). Among the 12 lines, only line~18 is for fixing the \textit{divide-by-zero} at line 19. The changes in \textit{getRMS} is code refactoring, i.e., the developer created \textit{getChiSquare}  to replace the deleted code at lines~3--7. Similarly, in \textit{getChiSquare} there is another refactoring that renames  variable \textit{sqHalf} to  \textit{chiSqHalf}. See lines~14, 15, 21, and 22.

In this example, the patch in \textit{getRMS} is completely not related to the vulnerability. If such patch is used as vulnerability detection data, we would introduce an incorrect label for this function. Even  for \textit{getChiSquare}, the majority lines in the commit are not related to the bug. Creating line-level vulnerability data using such commits can be too noisy. Similarly, uncleaned commits can also pose threats for other applications such as automatic program repair (APR). For example, the original patch for Math 65 from \textit{Defect4J}, caused CapGen~\cite{wen_context-aware_2018}, a state-of-the-art APR tool, to mislabel a correct patch it generated as incorrect due to not ``fixing'' the \textit{getRMS} function~\cite{yang_is_2021}. Furthermore, this also caused~\cite{sobreira_dissection_2018} to mislabel the type of changes required to fix the bug.




Our approach aims to use syntactic and semantic features of code commits and the code around the commits to distinguish whether a line is relevant to a bug. We hope to use machine learning models to learn patterns from existing commits which we know the ground truth labels, and use the models to predict whether a line in the commit is bug relevant or irrelevant. For example, for Figure~\ref{fig:CleanMath65}, the vulnerability fixing code (line~18), introduced an \textit{if} guard condition involving a non-zero check for a variable used within the \textit{if} block. Previous research~\cite{islam_how_2020, pan_toward_2009} have shown such changes are commonly used to fix vulnerabilities.

On the other hand, regarding the change of \textit{sqHalf} to \textit{chiSqHalf} at lines~14, 15, 21, and 22, we can see that there were no control flow changes. There were also no changes in data dependencies with other variables both within the patch and with respect to the entire function or file. These are commonly occurred patterns in refactoring. For refactoring in \textit{getRMS} function (lines~3--9), we can observe that the deleted variable \textit{criterion} was used inside a \textit{for} loop (lines~4--7), had some arithmetic operations and finally used as input to \textit{Math.sqrt} function at line~8. Meanwhile, after the patch, the changed input to \textit{Math.sqrt} is now a function (\textit{getChiSquare}). These type of patterns can help identify bug-irrelevant lines.

We trained an active learning model to classify each line within a commit as vulnerability-relevant or not based on the extracted features. We then used the model on function level vulnerability data and create the line-level vulnerability data.

%% file: overview.tex
\section{Approach}

\subsection{An Overview}\label{subsec:CleanFramework}
\begin{figure*}[ht]
  \centering
  \begin{subfigure}[b]{0.48\textwidth}
    \centering
    \includegraphics[scale=0.6]{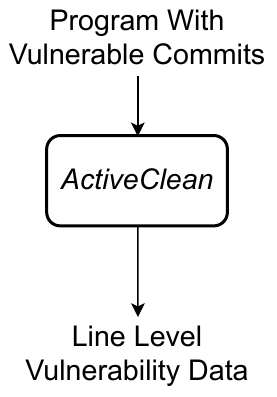}
    \includegraphics[scale=0.6]{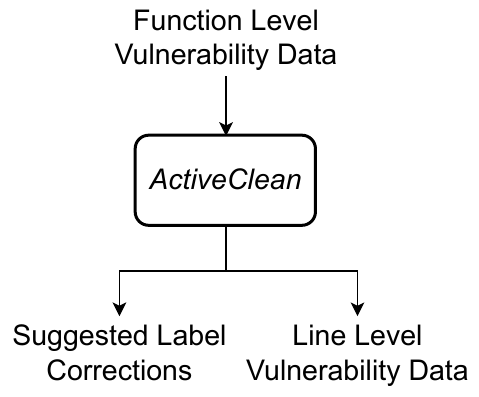}
    \caption{Inputs and Outputs for \tool{}}\label{subfig:Clean_Active_Input}
  \end{subfigure}
  \begin{subfigure}[b]{0.5\textwidth}
    \centering
    \includegraphics[scale=0.6]{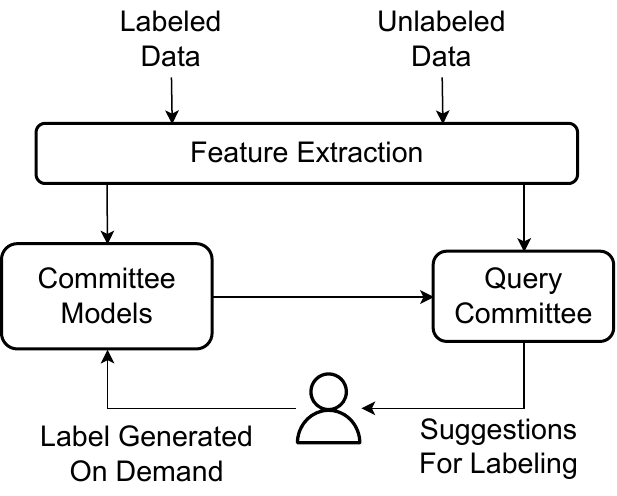}
    \caption{Training the Model using Active Learning}\label{subfig:Clean_Active_Train}
  \end{subfigure}
  \caption{Overview of \tool{}}\label{fig:CleanOverView}
\end{figure*}


Figure~\ref{fig:CleanOverView} presents an overview of our work. Figure~\ref{subfig:Clean_Active_Input} shows that the input of \tool{} is programs with their vulnerable commits. The output is line-level vulnerability data. \tool{} can also take input function level vulnerability data with commits. In that case, \tool{} will clean up the commits and generate line level labels, and meanwhile find mistakes that may exist in the function-level labels and suggest label corrections.

To achieve the goal, \tool{} trained a machine learning model via an active learning framework. See Figure~\ref{subfig:Clean_Active_Train}. The active learning framework used multiple machine learning models to form a \textit{committee}. The training starts with a small set of labeled data. The labeled data consists of a small amount of line-level vulnerability data. Their labels mark whether a line in the commit is relevant to the bug or not. Once the models are trained, the active learning starts. The committee instructs what commit lines from the (unlabeled) dataset we should label so that this dataset can achieve the best performance quickly. The user then generates labels for the commit lines recommended by the committee. These newly labeled data is used to train for improving the models. This process iterates until a predetermined budget, e.g., time, or the desired accuracy is reached. When applying to a new dataset, to further improve the performance, \tool{} can continue training with a small set of unlabeled data from the new dataset.

Once the active learning model is trained, we apply the final model on the input shown on Figure~\ref{subfig:Clean_Active_Input} to get line-level vulnerability data. If we observe a function contains an \textit{abnormal} amount of lines marked as non-vulnerable, we will report the function to the user for their further inspection and suggest that a non-vulnerable function may have been labeled as vulnerable. The user can configure the threshold to define what is \textit{abnormal} in their projects.




%% file: features.tex
\subsection{Feature Engineering}\label{subsec:CleanFeatureEng}
When manually cleaning a number of patches, code inspectors may not always be able to understand the detailed functionality of the code~\cite{herzig_impact_2013,yang_is_2021,croft2023data}. The patterns from syntactic and semantic code features may provide the clues to determine whether the code is relevant to the bug. For example, addition of conditional guards or \texttt{try-catch} blocks involving a variable from inside the block, like shown in Figure~\ref{fig:CleanMath65}, can indicate the presence of bug in the code within the wrapped block.

Our goal here is to apply machine learning to learn the patterns using the features from commits and the code context around the commit lines to automatically distinguish whether a commit line is relevant to the vulnerability. We studied 250 commits from a manually minimized Java dataset~\cite{herbold_fine-grained_2021} as well the past research~\cite{islam_how_2020, pan_toward_2009}, to identify a set of patterns and properties that may be useful for differentiating between vulnerability-relevant lines and vulnerability-irrelevant lines in the commits.



\begin{table*}[ht]
  \centering
  \resizebox{\textwidth}{!}{%
    \begin{tabular}
      {
      @{}llll@{}}
      \toprule
      \multicolumn{2}{c}{Commit Line}                             & \multicolumn{2}{c}{Commit Context}                          \\ \cmidrule(lr){1-2}\cmidrule(lr){3-4}
      Name            & Description                               & Name             & Description                              \\ \cmidrule(lr){1-2}\cmidrule(lr){3-4}
      assignment      & Contains assignment statement             & controlBlock     & Inside a control structure               \\
      comparator      & No. of comparator operators               & doBlock          & Inside a do-while block                  \\
      arithmetic      & No. of arithmetic operators               & ifBlock          & Inside an if block                       \\
      logical         & No. of logical operators                  & elseBlock        & Inside an else                           \\
      flagVar         & Sets/Removes boolean variable             & switchBlock      & Inside a switch block                    \\
      hasLiteral      & Contains literal strings                  & tryBlock         & Inside a try-catch                       \\
      isLocal         & Involves local variable                   & forBlock         & Inside a for blockblock block            \\
      hasRet          & Contain return statement                  & whileBlock       & Inside a while block                     \\
      funcCall        & Contain function call                     & dependedBy       & Total no. of control dependencies        \\ \cmidrule(l){1-2}
      \multicolumn{2}{c}{Within The Commit}                       & controlledBy     & Total no. of control dependencies        \\ \cmidrule(l){1-2}
      Feature         & Description                               & reachableOutside & Has data dependency outside commit lines \\\cmidrule(lr){1-2}
      controlDepend   & Control dependent on other commit lines   & postDomintaedBy  & No. of lines it is post dominated by     \\
      depends         & Data dependent on other commit lines      & postDominates    & No. of lines it post dominates           \\
      repeated        & No. of repetition within the commit       &                  &                                          \\
      repeatedCall    & No. of same function call repetitions     &                  &                                          \\
      repeatedControl & No. of same control structure repetitions &                  &                                          \\ \bottomrule
    \end{tabular}%
  }
  \caption{Features Designed For Identifying Vulnerability-relevant Lines}\label{tab:CleanFeatures}
\end{table*}

In Table~\ref{tab:CleanFeatures}, we present the features we designed. The features can be classified into three categories; (1) features that capture characteristics of each individual commit lines themselves (shown under \textit{Commit Lines}), (2) features that capture how different commit lines interact with each other (under \textit{Within The Commit}), and finally (3) features that capture how the commit lines interact with rest of the code (under \textit{Commit Context}).

The intuition behind the \textit{Commit Lines} features was to identify simple buggy patterns that may exist at line level. For example, the \textit{comparator}, \textit{arithmetic} and \textit{logical} features shown in Table~\ref{tab:CleanFeatures} under \textit{Commit Line} capture the number of each operator type present in a commit line. The often occurring off-by-one errors are fixed by changing the type or numbers of one of these operator and can be captured by them. As another example, common bug fixes typically involve ``returning from the function and reporting an error message''. This can be captured using \textit{hasRet}, which checks if the line contains
a {\tt return statement}, in conjunction with \textit{hasLiteral}, which check for strings in the statements. Similarly, using a \texttt{Boolean} variable to avoid a buggy control flow can be captured using the \textit{flagVar} feature.

The features in \textit{Within The Commit} were designed to differentiate the types of changes of \textit{functionality enhancement} and \textit{refactoring} from bug fixes. For example, most bug fixes only change their control and data dependencies minimally. On the other hand, functionality enhancement and refactorings introduce substantially more control and data dependencies both within the commit and around it. This can be captured using the change to \textit{controlDepend} and \textit{depends} features (see Table~\ref{tab:CleanFeatures}), which reports the number of control and data dependency respectively. Features like \textit{repeated}, \textit{repeatedCall}, and \textit{repeatedControl} are designed to capture the patterns commonly found in refactoring changes where the same code or function calls are repeated throughout the commits.

\textit{Commit Context} captures the location of the commit lines as well as how they interact with the program in general as they provide valuable clues to determining if the code is relevant to the bug. This is based on our observation that most bug fixes tend to happen within existing control, functional, or loop body. The \textit{controlBlock} and other block based features (see Table~\ref{tab:CleanFeatures}) captures whether the change is happening inside its namesake's body or not. Functionality enhancement and refactorings changes, on the other hand, frequently introduce new functions and variables. These newly introduced functions and variables have a higher number of control and data dependencies among themselves than with the context surrounding it. The block based features along with \textit{dependedBy}, \textit{controlledBy} and other features check for control and data dependency with the surrounding context to capture this.

We aim to design features that were independent of the underlying implementation language and project. Hence, we don't use any actual code to avoid learning project specific function or variable names and used general semantic patterns instead of more detailed language dependent ones. We also designed them to be easy for machine learning consumption by only using boolean and number based features.

In this work we use \textit{code property graphs (CPG)} that integrates control flow graph, abstract syntax tree and dependency graph~\cite{yamaguchi_modeling_2014} to represent source code from commits and extract the features. Specifically, we use its implementation in Joern~\cite{joernio}, a code analysis platform, which supports multiple languages including C, C++, Java and Python.

In order to extract features, for each commit we retrieved both the current and before the patch versions of all the source code files present in a commit. Then we used Joern to build CPGs for them. Next, we analyzed the syntactic and semantic properties for each individual commit lines using Joern and extracted the features listed in Table~\ref{tab:CleanFeatures} with custom CPG queries we designed.




%% file: learning.tex
\subsection{Applying Active Learning for Cleaning the Commits}\label{sec:CleanActiveLearn}
We formulate the patch cleaning problem to a machine learning classification problem. Here the input is the vector that represents the features of a commit line. The output is a label, ``1'' indicates that the line is relevant to the vulnerability; ``0'' indicates that the line is other types of changes, e.g., refactoring or function enhancement.

We used \textit{active learning}, specifically, the \textit{query-by-committee} approach, to learn the patterns that identify bug relevant commit lines from the extracted features. \textit{Query-by-committee} is a very effective active learning approach that has been successfully applied to different classification problems~\cite{melville_diverse_2004}. Specifically, we used the \textit{query-by-committee} using \textit{vote-entropy} as the disagreement metric from modAL~\cite{modAL2018}, a machine learning framework built on top of Scikit-learn~\cite{scikit-learn}.

A \textit{committee} can consist of two or more traditional machine learning approaches like Random Forest or Support Vector Machines. Then given a set of unlabeled data, \textit{query-by-committee} iteratively selects data to be labeled for training. Typically, the selection is determined by some measure of disagreement in the committee about its predicted labels. The committee typically selects the examples with the least certainty and maximum disagreement, as labeling such examples bring in the maximum information to the models. Using this approach, active learning typically required much fewer labels and can quickly learn a good model.


In our approach, first, we take the extracted features along with known line-level ground truth labels to train a set of initial base models for the committee. Next, we extract features from commit lines without line-level ground truth data. We then use the disagreement between the predictions from the different \textit{committee} models to select a fixed amount of additional data to be labeled in each iteration. A user manually inspects and provide labels for these commit lines. Finally, we repeat the querying and labeling until a fixed budget, e.g., measured by time or the numbers of data we can label, is exhausted.


If the input to the model is program with vulnerable commits (shown in Figure~\ref{subfig:Clean_Active_Input}), we used the final model after active learning to predict if each individual commit line is relevant to the bug or not. Based on the prediction for each line, the commit lines are encoded as ``1'' or ``0'' depending on if they were relevant to the vulnerability or not, respectively, as the output.

For function level vulnerability data, on top of line-level data, we also output the line numbers of the vulnerable lines from the function and whether a vulnerable function needs to be reexamined due to abnormal amount of non-vulnerable lines as the output.





%% file: results.tex
\section{Evaluation}\label{sec:CleanEvaluation}
In this paper, we aim to answer the following research questions:

\begin{itemize}
  \item \textbf{
          RQ1 [Validation]} Can our approach effectively and efficiently clean the buggy commits?
  \item \textbf{
          RQ2 [Comparison]} How does \tool{} compared to other baselines and settings?
  \item \textbf{
          RQ3 [Application]} How our approach can help deep learning based vulnerability detection?
\end{itemize}

\subsection{Experimental Setup}\label{subsec:CleanExprSetup}
\subsubsection{Implementation} We implemented \tool{} for C and Java programs using Joern~\cite{joernio}, Scikit-learn~\cite{scikit-learn}, modAL~\cite{modAL2018}, Python, Bash, and Scala. Specifically, we used Joern with Bash scripts and Scala to extract features from patches. Then we used Scikit-learn along with modAL for active learning.

\subsubsection{Subject selection}\label{subsubsec:CleanSelectSub} To answer the research questions and demonstrate that our techniques are applicable in practice, we aim to use benchmarks that (1) are real-world open source programs for Java and C, (2) have manually labelled or verified bug fixing commits, so we can have ground truth to compare against, (3) and are actively used in the research community, so our results can directly benefit the users.


We searched for readily available benchmarks based on the above criteria in the literature~\cite{herbold_fine-grained_2021,jiang_bugbuilder_2023,wang_cora_2019,zheng_d2a_2021,chakraborty_deep_2020,zhou_devign_2019,kirinuki_hey_2014,yang_is_2021,herzig_impact_2013,xu_tracking_2022,chen_untangling_2022,dias_untangling_2015,just_defects4j_2014,do_supporting_2005,lu_bugbench_nodate,le_goues_manybugs_2015,bohme_where_2017,dallmeier_extraction_2007}. As a result, we choose all the 17 Apache projects in~\cite{herbold_fine-grained_2021} for Java programs, as it provided manually verified line-level ground truth labels for bug fixing commits. This gave us 27K line-level labels spread over 365 commits. We were however, unable to find an equivalent benchmark for C programs. While benchmarks like SIR~\cite{do_supporting_2005}, ManyBugs~\cite{le_goues_manybugs_2015} and DBGBench~\cite{bohme_where_2017} were promising, SIR only contains seeded bugs. ManyBugs and DBGBench reported very few bugs per program.

Thus, we used FFmpeg and QEMU from Devign~\cite{zhou_devign_2019} in our study. The dataset contain function level vulnerability labels and are widely cited. Two authors of this paper manually inspected the commits and provided line-level labels for a small set of data as needed by our evaluation, following the manual inspection protocol documented in the literature~\cite{islam_comprehensive_2019}. Given a budget of 21 days, the two authors provided labels for 260 (170 FFmpeg + 100 QEMU) randomly selected commits. There were disagreement over 13 (3 FFmpeg + 10 QEMU) of the 260 commits' labels even after discussions between the authors. Hence, we excluded them from our dataset. In total, we generated line-labels for 4.4K (3.2K FFmpeg + 1.2K QEMU) lines over 257 (167 FFmpeg + 90 QEMU) commits for C code.

\subsubsection{Selecting Machine Learning Models to Set up the Active Learning Framework}\label{subsubsec:CleanSelecML} We apply the \textit{query-by-committee}~\cite{modAL2018} based active learning framework to train a model. The committee consisted of two models. To select the models, we considered five well-known machine learning models: namely \textit{Random Forest}~\cite{breiman_randomF_2001}, \textit{Label Spreading}~\cite{zhou_learning_nodate}, \textit{Label Propagation}~\cite{zhu_learningFL_2002}, \textit{Support Vector Machine}~\cite{cortes_svm_1995}, and \textit{Logistic Regression}~\cite{cox_regression_1958}. We used the ground truth labels from \textit{spike} and \textit{giraph}, two other Java projects from~\cite{herbold_fine-grained_2021}, to select the \textit{committee}. Based on this initial analysis, \tool{} was implemented as a \textit{query-by-committee} active learning model using \textit{Random Forest} and \textit{Support Vector Machine}.

\subsubsection{Experimental design for RQ1.}\label{subsubsec:CleanExpSetupRq1} In RQ1, our goal is to evaluate the effectiveness and efficiency of the of labels generated using \tool{}. We measured \textit{effectiveness} by evaluating the reduction of commits and the correctness of the reduction. Specifically, for reduction of commits we measure (1) the average number of bug-irrelevant lines removed, and (2) the number of commits that had their bug-irrelevant lines removed. For the correctness of the reduction, we reported the F1 score for the prediction. For \textit{efficiency}, we report the number of training examples needed against the F1 score achieved via \tool{}.


For the Java dataset, we used the line-level ground truth labels provided by the benchmark~\cite{herbold_fine-grained_2021} to evaluate the effectiveness and efficiency of our approach using 3-fold cross validation. Our setup follows the active learning literature~\cite{herzig_impact_2013,yu_improving_2021}. First, we trained \tool{} using 20\% of the training data to set up the base models in the committee. In the second step, we queried the committee for 200 iterations and 10 commit lines per interaction from the remaining 80\% of the training data. After training the additional 2K commit lines, we plotted the F1 score at query intervals to evaluate the learning efficiency. Finally, we used the prediction made on the testing data to collect the effectiveness metrics.

To train base models for C projects, we used the entire Java dataset. This is because we only can label a small set of C data and they were all used as test data. In the second step, we queried the committee for 100 iterations with 10 commit lines (FFmpeg) per iteration. The labels for these 1K commit lines were generated on demand by one author. After the training, we plotted the F1 score at query intervals to evaluate the improvement in efficiency. Finally, for evaluating effectiveness, we used the prediction made on the testing data as well as the remaining unlabeled FFmpeg data. Since we don't have ground truth for unlabeled FFmpeg data, we randomly selected 500 commit lines (250 each from vulnerability fixing and bug-irrelevant predictions) and manually evaluated them using our manual inspection protocol given above.


Finally, we wanted to evaluate the effectiveness of \tool{} for new projects. Specifically, we wanted to measure how \tool{} performed on an unseen C project and whether we can quickly adapt \tool{} for new projects. For this experiment, we used QEMU, another C dataset present in Devign. Two authors manually inspected and labeled 100 commits using the manual inspection protocol given above. Due to disagreement, we threw out 10 commits' labels. We created 2.8 K labels and used this labeled dataset as the test set. We then apply \tool{} that was trained on Java and FFmpeg dataset from above to the QEMU test set. The performance was measured using the F1 score. Next, in order to measure adaptability, we used the query-by-committee to query and teach 100 more labels. We used the change in F1 score to measure the adaptability.

\subsubsection{Experimental design for RQ2.}\label{subsubsec:CleanExpSetupRq2}
To compare the active learning approach we used, we selected top 3 machine learning models as baselines, namely \textit{Label Spreading}, \textit{Random Forest} and \textit{Support Vector Machine}. The models are chosen because they rank top for the \textit{spike} and \textit{giraph} datasets from Section~\ref{subsubsec:CleanSelecML}. We trained each model 200 times, starting with 20\% of the training data and then adding 10 more training data each time. That helps us understand what are the minimum data needed for those models to achieve the top performance, so that we can compare them with active learning. We have also searched the literature for other baselines~\cite{jiang_bugbuilder_2023, chen_untangling_2022, cao_mvd_2022, shen_smartcommit_2021, partachi_flexeme_2020, wang_cora_2019, dias_untangling_2015, barnett_helping_2015, kirinuki_splitting_2016, yamashita_changebeadsthreader_2020, fu_linevul_2022} that can clean line-level commits. Among the tools we are able to run, (1) \textit{FLEXME} handled only \texttt{C\#} projects; (2) \textit{LineVul}~\cite{fu_linevul_2022} is able to predict line-level vulnerability, but it reported less than 15\% F1 due to our benchmarks being out of distribution.


In RQ2, we also performed two ablation studies. In the first study, we aim to study the impact of model selection for \textit{query committee}. For this, we trained 3 models with different \textit{query committee}. Specifically, we used \textit{query committee} with \textit{Random Forest} in conjunction with \textit{Logistic Regression}, \textit{Label Spreading}, and \textit{Label Propagation}, as these gave the best performance in our tuning dataset (see Section~\ref{subsubsec:CleanSelecML}). These models were trained using the same initial training data as \tool{}. Then we used their \textit{committees} to select the same amount of additional training data as \tool{} from the rest of the training dataset.

In the second study, we aim to understand the advantage of applying active learning compared to random sampling of the labeled data. Hence, we used the same base models trained with 20\% training data. Instead of using active learning, we randomly sampled the same amount of data at each iteration the same way we train \tool{}. We plot the change of F1 scores over iterations for the two approaches. For both the baseline and ablation experiments, we used the Java dataset to ensure that we can use a large amount of labeled data.

\subsubsection{Experimental design for RQ3.}\label{subsubsec:CleanExpSetupRq3} RQ3 aims to demonstrate the usefulness of \tool{}. For our first application, we apply our cleaned dataset in the line-level vulnerability detection and compare the model performance with and without our cleaned data. We used LineVul~\cite{fu_linevul_2022}, the SOTA and only line-level vulnerability detection model that is able to run on FFmpeg. We trained LineVul with 9.5K functions of FFmpeg. In the vanilla setting, we use commits of the functions as line-level labels, and in the \tool{} setting, we use our cleaned dataset. To compare test set performance, we used the metrics of F1 score, the top 10 accuracy and \textit{IFA} (the average position of the first vulnerable line within the top 10) to measure the accuracy of predicted vulnerable lines compared to the ground truth.

In the second application, we explored the possibility of using \tool{} to suggest label corrections. Specifically, we recommended any vulnerable function with more than 50\% vulnerability-irrelevant lines for reclassification. We randomly sampled 50 functions which \tool{} proposed as vulnerability-irrelevant, and then applied the manual inspection to determine their correctness.

Since \tool{} can correct function-level labels, we also ran LineVul for function level prediction using the labels from Devign and using the labels from \tool{} respectively. We used F1 scores to compare the performance for the two settings.

For the experiments with LineVul, we used the same test sets for baseline and \tool{}. They are the manually labeled ground truth from FFmpeg.

\subsubsection{Running the experiments.} The feature extraction, \tool{} and baseline models were trained and evaluated on a VM with 64-bit 16 core Intel Haswell processor and 32 GB of RAM\@. The LineVul models were evaluated on a VM with a 32 core CPU and GPU with 16 GB RAM\@. Both the VMs were running CentOS 8.

\subsection{Results for RQ1: Validation}\label{subsec:CleanRq1Result}
\begin{table*}[ht]
  \centering
  \begin{tabular}
    {
    @{}l|llll@{}}\toprule
    Dataset                                              & Java test & FFmpeg test & FFmpeg application & Qemu new project test \\ \midrule \midrule
    \tool{} F1 Score                                     & 70.23     & 74.83       & ---                & 64.46                 \\
    Total Commit Lines                                   & 8909      & 3129        & 98267              & 1200                  \\
    Total Bug-irrelevant Lines (Predicted)               & 6832      & 1202        & 51950              & 748                   \\
    Total Bug-irrelevant Lines (Confirmed)               & 5792      & 544         & 185/249            & 419                   \\ \midrule
    Total Commits                                        & 364       & 167         & 5246               & 90                    \\
    Commits With Bug-irrelevant Lines (Predicted)        & 342       & 114         & 3628               & 76                    \\
    Commits With Bug-irrelevant Lines (Confirmed)        & 332       & 57          & ---                & 43                    \\ \midrule
    Average Size Of Commits                              & 25.09     & 18.73       & 17.01              & 13.33                 \\
    Average Bug-irrelevant Lines Per Commits (Predicted) & 20.42     & 10.54       & 14.31              & 9.84                  \\
    Average Bug-irrelevant Lines Per Commits (Confirmed) & 17.49     & 6.03        & ---                & 9.74                  \\ \bottomrule
  \end{tabular}%
  \caption{Results for RQ1: Effectiveness}\label{tab:CleanRq1Efct}
\end{table*}
\begin{figure*}[ht]
  \centering
  \begin{subfigure}[b]{\columnwidth}
    \centering
    \includegraphics[scale=0.5]{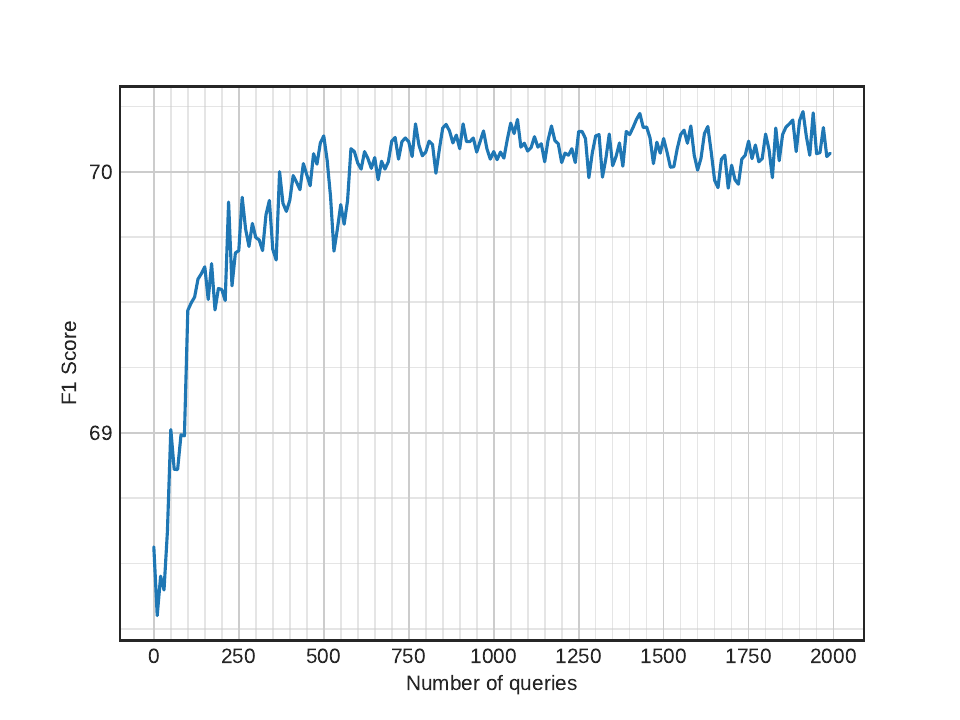}
    \caption{F1 Score vs number of labeled data for \textit{Java test} for 2K additional training data}\label{fig:Clean_f1_java}
  \end{subfigure}
  \hfill
  \begin{subfigure}[b]{\columnwidth}
    \centering
    \includegraphics[scale=0.5]{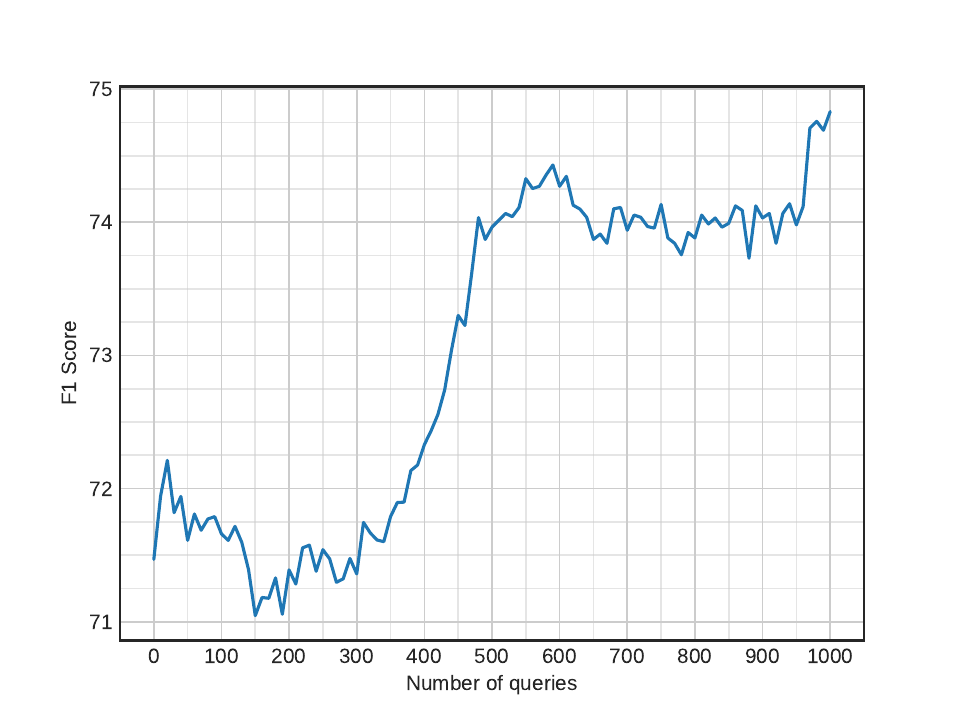}
    \caption{F1 Score vs number of labeled data for \textit{FFmpeg test} for 1K additional training data}\label{fig:Clean_f1_1k_c}
  \end{subfigure}
  \caption{Results for RQ1: Efficiency measured as F1 score against number of trained labeled data}\label{fig:CleanRq1Eff}
\end{figure*}


In Table~\ref{tab:CleanRq1Efct}, we present the results for the Java dataset labeled from~\cite{herbold_fine-grained_2021}, the manually labeled FFmpeg, and Qemu, under the columns \textit{Java test}, \textit{FFmpeg test} and \textit{Qemu new project test} respectively. Under \textit{FFmpeg application}, we report the results of processing the entire FFmpeg dataset (excluding the manually labeled ones under \textit{FFmpeg test}) in Devign.


As shown in row \textit{\tool{} F1 Score}, \tool{} reported the highest F1 score of 74.83 for \textit{FFmpeg test}, and it reported with F1 score 70.23 for  \textit{Java test}. Since we don't have the line-level labels for \textit{FFmpeg application}, we did not report F1; however, we sampled results to manually confirm the predictions (see later). When applying to \textit{Qemu} without any retraining, we report F1 score of 59.53. After providing 100 line additional labels, we obtained 64.46 F1 shown under \textit{Qemu new project test}.




\tool{} is able to reduce bug irrelevant commit lines across all the datasets, as shown in the rows \textit{Total Bug-irrelevant Lines (Predicted)} and \textit{Total Bug-irrelevant Lines (Confirmed)}. For \textit{FFmpeg application}, we sampled 500 data for inspection (250 bug-relevant, 250 bug-irrelevant). Among the 250 bug irrelevant lines reported by \tool{}, we confirmed that 185 were correctly predicted, while among bug relevant lines, we confirmed that 187 out of 250 were indeed bug relevant. We excluded 8  (one from irrelevant and 7 from relevant) out of 500 for consideration because the two authors were not able to achieve an agreement. The significantly higher number of bug-irrelevant lines in \textit{Java test} were due to documentation and test cases changes that were present in bug fixing commits for this dataset.

\tool{} also showed that the bug-irrelevant lines were spread across commits in all the datasets, as shown in the rows \textit{Commits With Bug-irrelevant Lines (Predicted)} and \textit{Commits With Bug-irrelevant Lines (Confirmed)}. Thus, the line-level cleaning tool like \tool{} can be useful for many commits. Even for the Devign datasets of \textit{FFmpeg test} and \textit{Qemu new project test} that likely have been cleaned out documentation and test (unlike \textit{Java test}) but focus on only the vulnerability functions, we reported many commits needed to be cleaned.


The prevalence of bug-irrelevant lines per commit is shown in rows \textit{Average Bug-irrelevant Lines Per Commit (Predicted)} and \textit{Average Bug-irrelevant Lines Per Commit (Confirmed)}, while the average size of the commits is shown in \textit{Average Size Of Commit}. For the \textit{Java test} dataset, on average 17.5 lines out of a commit with 25 lines were bug-irrelevant. Interestingly, \textit{Qemu new project test} reported an average of more than 9 bug-irrelevant lines per commit when the average size was only around 13. This indicates that on average, there are a large portion of commits that need to be cleaned out as they are not related to vulnerability. Directly using commits as line-level labels can bring in much noise for line-level vulnerability detection.

The Figure~\ref{fig:CleanRq1Eff} shows the results for efficiency of \tool{}.
Y axis is F1 score. X axis plots the number of labeled lines provided during the period of active learning.
Figure~\ref{fig:Clean_f1_java} and Figure~\ref{fig:Clean_f1_1k_c} show the corresponding plots for \textit{Java test} and \textit{FFmpeg test} respectively.


The results show that \tool{} is able to quickly improve the F1 scores for both datasets. Figure~\ref{fig:Clean_f1_java} shows that only 400 additional commit lines were needed for \tool{} to reach an F1 score of 70 for \textit{Java test}. Similarly, an additional 400 training data improved the F1 score for \textit{FFmpeg test} to 74 from 71.5 as shown in Figure~\ref{fig:Clean_f1_1k_c}. To get a complete picture of the efficiency, we trained a model with the entire training data available (a total of 6.6K labels) for \textit{Java test}. This model had an F1 score of 72.82. That said, \tool{} only needed 400 of data (on top of base models trained with 20\%) to reach close to the best performance.

\subsection{Results for RQ2: Comparison}\label{subsec:CleanRq2Result}
\begin{table}[ht]
  \centering
  \begin{tabular}
    {
    @{}ll@{}} \toprule
    Baseline Models        & Average F1      \\ \midrule
    \tool{}                & \textbf{70.230} \\
    Label Spreading        & 67.187          \\
    Random Forest          & 67.175          \\
    Support Vector Machine & 67.076          \\ \bottomrule
  \end{tabular}%
  \caption{Results for RQ2: Baseline Comparison: Average F1 scores after final training}\label{tab:CleanRq2F1Full_Baseline}
\end{table}
\vspace{-15pt}

Table~\ref{tab:CleanRq2F1Full_Baseline} shows the average F1 scores for the different baselines after the final training. \tool{} got the best average F1 score of 70.23. The baselines all performed similarly at 67.18, 67.17, and 67.07 for \textit{Label Spreading}, \textit{Random Forest}, and \textit{Support Vector Machine}, respectively.

\begin{figure}[ht]
  \centering
  \includegraphics[scale=0.5]{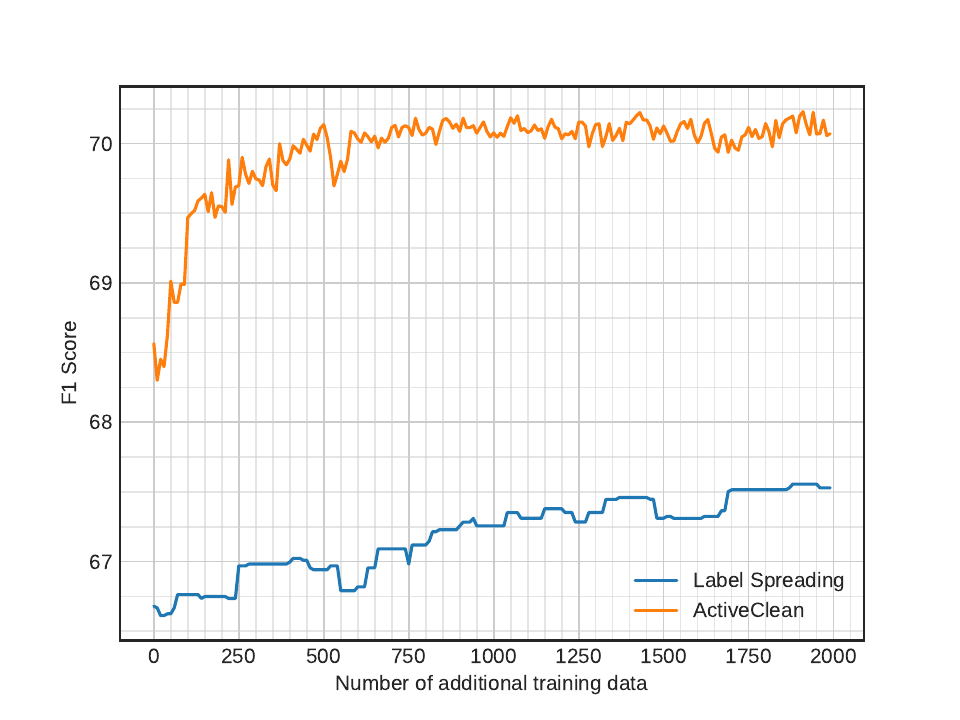}
  \caption{Results for RQ2: Baseline Comparison: Efficiency of Label Spreading (the best baseline) compared to \tool{}}\label{fig:CleanRq2R_LSpread_Rand}
\end{figure}

In Figure~\ref{fig:CleanRq2R_LSpread_Rand}, we have plotted the F1 score at different amount of training data for \textit{Label Spreading} (shown in blue), which was the best baseline. We also plot the F1 score for \tool{} (shown in orange), for comparison. As we can see, just after the initial training, the \textit{committee} based \tool{} reached an F1 score of 68.56 compared to the \textit{Label Spreading} model's 66.68. This initial lead, is further improved quickly by learning the data suggested by querying the committee. \tool{} took just 50 additional data to get to 69 and 400 for reaching 70. Meanwhile, even after 2K additional training data, \textit{Label Spreading} was unable to improve even 1 full point. This clearly demonstrated the advantage of active learning that it can learn very quickly with lesser training data.



We present the results for the two ablation study in Table~\ref{tab:CleanRq2F1Full_Albation}. In the first study where we tried active learning using different \textit{committees}. \tool{} leads with an F1 score of 70.23 followed closely by the model with \textit{Random Forest + Logistic Regression} at 69.10. The models with \textit{Label Propagation} and \textit{Label Spreading} performed similarly at 68.812 and 68.765. However, all the different \textit{committee} based models performed better than the baseline models.

\begin{table}[ht]
  \centering
  \begin{tabular}
    {
    @{}ll@{}} \toprule
    Models                                 & Average F1      \\ \midrule
    \tool{}                                & \textbf{70.230} \\ \midrule
    \multicolumn{2}{c}{Ablation Study: Active Learning}      \\ \midrule
    Random Forest + Logistic Regression    & 69.108          \\
    Random Forest + Label Propagation      & 68.812          \\
    Random Forest + Label Spreading        & 68.765          \\ \midrule
    \multicolumn{2}{c}{Ablation Study: Random Selection}     \\ \midrule
    Random Forest + Support Vector Machine & 69.031          \\ \bottomrule
  \end{tabular}%
  \caption{Results for RQ2: Ablation Studies: Average F1 scores after final training}\label{tab:CleanRq2F1Full_Albation}
\end{table}
\vspace{-15pt}

\begin{figure*}[ht]
  \centering
  \begin{subfigure}[b]{\columnwidth}
    \centering
    \includegraphics[scale=0.5]{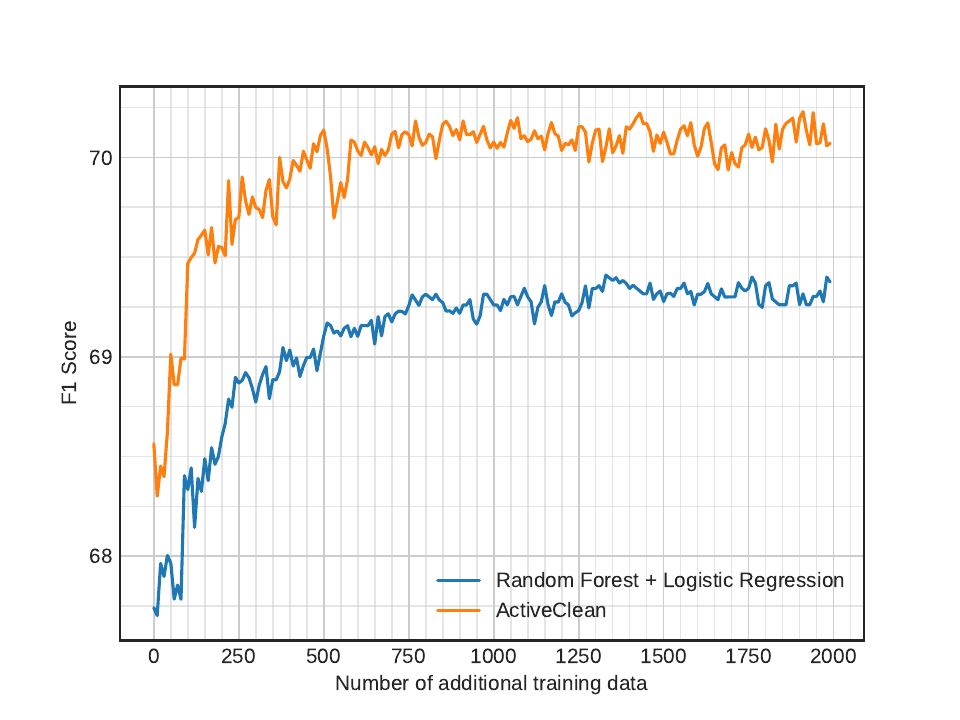}
    \caption{Active Learning Ablation Study: Random Forest + Logistic Regression (the second best setting) compared to \tool{}}\label{subfig:CleanRq2R_LRegr}
  \end{subfigure}
  \hfill
  \begin{subfigure}[b]{\columnwidth}
    \centering
    \includegraphics[scale=0.5]{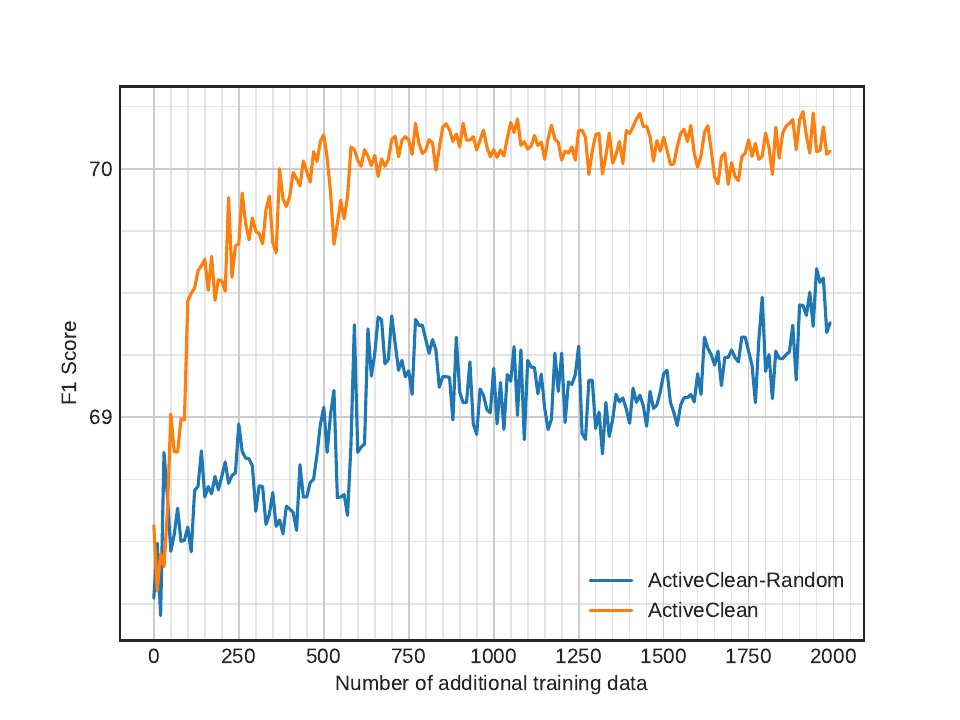}
    \caption{Random Learning Ablation Study: Random Selection based \tool{} compared to \tool{}}\label{subfig:CleanRq2R_SVM_Rand}
  \end{subfigure}
  \caption{Results for RQ2: Efficiency measured as F1 vs amount of additional training data}\label{fig:CleanRq2_F1_Eff}
\end{figure*}

The efficiency comparison for this study is shown in Figure~\ref{subfig:CleanRq2R_LRegr}. The F1 score at different amount of additional training data for \tool{} is shown in orange and \textit{Random Forest + Logistic Regression}, the next best setting, is shown in blue. After the initial training, \tool{} had an F1 of 68.54 while \textit{Random Forest + Logistic Regression} had 67.74. \tool{}, improved the F1 score by 1 point to 69.58 after training 120 additional commit lines. Meanwhile, it took \textit{Random Forest + Logistic Regression} 240, double, the additional training data to improve 1 point to 68.74. However, in general all the \textit{committee} based models shows steady improvements, but at slower growth curves. For example, model with \textit{Random Forest + Label Propagation} took 1K additional training data to reach an F1 of 69.

For the second ablation study using random selection, the model had an F1 score of 69.03 at the end of 2K additional training data compared to \tool{}'s F1 score 70.23. We found that using random selection with \tool{}'s \textit{committee} was still better than other selections of the \textit{committees}. Figure~\ref{subfig:CleanRq2R_SVM_Rand}, shows the efficiency comparison where \tool{} (in orange) showed faster improvement in F1 score per additional training data than the same \textit{committee} when trained with random selection (shown in blue as \tool{}-Random). Both models started out at 68.54 F1 score after initial training. But it took \tool{} only 50 additional training data to pass an F1 score of 69, while \tool{}-Random took 500. While \tool{} finally reached an F1 score of 70.21, \tool{}-Random only reached 69.341 with 2K addition data.

\subsection{Results for RQ3: Applications}\label{subsec:CleanRq3Results}
\begin{table}[ht]
  \begin{tabular}
    {
    @{}lccccc@{}} \toprule
    Model          & F1          & Top 10        & IFA           & \multicolumn{2}{c}{Correctly Predicted} \\ \midrule
                   &             &               &               & Functions    & Lines                    \\ \cmidrule(l){5-6}
    \tool{}-FFmpeg & \textbf{87} & \textbf{73\%} & \textbf{6.68} & \textbf{238} & \textbf{178}             \\
    FFmpeg         & 83          & 66\%          & 8.11          & 220          & 108                      \\ \bottomrule
  \end{tabular}
  \caption{Results for RQ3: LineVul Ran with Cleaned Dataset}\label{tab:CleanRq3LineVul}
\end{table}

At the line level, using the clean dataset \textit{\tool{}-FFmpeg}, LineVul reported 178 vulnerable lines while the vanilla dataset~\textit{FFmpeg} only reported 108 vulnerable lines. This is a 1.6 X improvement resulting in 70 more vulnerable lines. The \textit{\tool{}-FFmpeg} model also reported 73\% \textit{Top 10} accuracy, compared to 66\% reported by \textit{FFmpeg} model. The \textit{\tool{}-FFmpeg} model reported an IFA of 6.68 as opposed to 8.11, an improvement of 1.43. Lower IFA means the lesser number of bug-irrelevant lines within the Top 10 ranking of the model. In another word, a developer would typically have to inspect 1.5 lines less per function in order to reach the vulnerable line.

\begin{table}[ht]
  \centering
  \begin{tabular}
    {
    @{}ll@{}} \toprule
    Number of                & Total \\ \midrule
    Functions Examined       & 50    \\
    Non-Vulnerable Functions & 29    \\
    Vulnerable Functions     & 13    \\
    Undecided                & 8     \\ \bottomrule
  \end{tabular}%
  \caption{Results for RQ3: Help Correct Function-Level Labels}\label{tab:CleanRq3Reclass}
\end{table}
\vspace{-15pt}

In Table~\ref{tab:CleanRq3Reclass}, we show that \tool{} can help correct function-level labels. In our experiments, \tool{} recommended 468 label correction for the FFmpeg dataset. We randomly sampled 50 functions for inspection, and the two authors agreed that 29 functions or 69.04\% were indeed non-vulnerable.  

%% file: examples.tex
\subsection{Interesting Examples}
In this section we give two interesting examples that \tool{} recommended as incorrect function-level labels.

\begin{figure}[ht]
  \centering
\begin{lstlisting}[language=diff]
+ static av_cold int alac_decode_close(AVCodecContext *avctx)  {
+     ALACContext *alac = avctx->priv_data;
+   int chan;
+   for (chan = 0; chan < alac->numchannels; chan++) {
+       av_freep(&alac->predicterror_buffer[chan]);
+     ...
+     return 0;
+ }
 static av_cold int alac_decode_init(AVCodecContext * avctx) {
    int ret;
    ALACContext *alac = avctx->priv_data;
    alac->avctx = avctx;
    ...
    allocate_buffers(alac);
-   return 0;
- }
- static av_cold int alac_decode_close(AVCodecContext *avctx) {
-     ALACContext *alac = avctx->priv_data;
-   int chan;
-   for (chan = 0; chan < alac->numchannels; chan++) {
-       av_freep(&alac->predicterror_buffer[chan]);
-     ...
+   if ((ret = allocate_buffers(alac)) < 0) {
+       av_log(avctx, AV_LOG_ERROR, "Error allocating buffers\n");
+       return ret;
    return 0;
  }
\end{lstlisting}
  \caption{an incorrect label in FFmpeg in the Devign dataset}\label{fig:CleanRefactor}
\end{figure}

In Figure~\ref{fig:CleanRefactor}, we show an example in \texttt{FFmpeg} where Devign provided an incorrect function-level label for \texttt{alac\_decode\_close} (the older version). The diff patch~\footnote{\url{https://github.com/FFmpeg/FFmpeg/commit/53df079a730043cd0aa330c9aba7950034b1424f}} is shown in Figure~\ref{fig:CleanRefactor}, where red indicates deletion and green indicates addition. Here, lines 17--22 from the older function were moved to lines 1--6. This function is not vulnerable and there is no correction added for this function in the commits. \tool{} correctly detected this. 

\begin{figure}[ht]
  \centering
\begin{lstlisting}[language=diff]
 static inline void RENAME(yuvPlanartoyuy2)(const uint8_t *ysrc, const uint8_t *usrc, const uint8_t *vsrc, uint8_t *dst,
-  long width, long height,
-  long lumStride, long chromStride, long dstStride, long vertLumPerChroma)
+            long width, long height,
+            long lumStride, long chromStride, long dstStride, long vertLumPerChroma)
 {
-  long y;
-  const long chromWidth= width>>1;
-  for(y=0; y<height; y++)
-  {
+    long y;
+    const long chromWidth= width>>1;
+    for (y=0; y<height; y++)
+    {
\end{lstlisting}
  \caption{Mislabeled Formatting Changes as Vulnerable}\label{fig:CleanExSpace}
\end{figure}
Figure~\ref{fig:CleanExSpace} shows a patch~\footnote{\url{https://github.com/FFmpeg/FFmpeg/commit/6e42e6c4b410dbef8b593c2d796a5dad95f89ee4}} that only formats the code, but the function \texttt{RENAME} at line~1 was labeled as vulnerable in the Devign dataset. \tool{} recommended to correct this label. Interestingly, this patch changed 5 files with over 6K changes, all of which are formatting changes. Most of the functions inside this commit was recommended for label correction by \tool{}.

%% file: threats.tex
\section{Threats to Validity}\label{sec:CleanThreat}

\noindent{\bf Internal Threats to Validity:} One of the important challenges we face is the need of ground truth for cleaned commits. Hence, we used the dataset provided by~\cite{herbold_fine-grained_2021} for our Java projects, which required four different authors to review the labels with at least 3 authors agreeing with the labels. The ground truth generated for our C Project used two author agreement following~\cite{islam_comprehensive_2019}. To ensure there is no model parameter selection bias, we used entirely different projects for the model selection than those used for in the experiments. We also used 3-fold cross validation for our experiments to ensure that are results doesn't show selection bias. We also sampled the results that did not have ground truth and manually validated them using two authors.


\noindent{\bf External Threats to Validity:} To mitigate external threats, we used ground truth labels from two datasets containing Java and C real-world open-source projects covering more than 4.3K commits with 119K commit lines. These projects have very different purpose, size, and authors to ensure that \tool{} doesn't learn and report project specific traits. We have evaluated \tool{} on 17 java projects and 2 C real-world repositories from widely used the vulnerability dataset Devign. To the best of our knowledge, there are LineVul, LineVD, and IVDetect three line-level vulnerability detection tools. Only LineVul can be run successfully, and it is the SOTA.




%% file: related.tex
\section{Related Work}


Several studies have emphasized the need for large high quality data for vulnerability detection tasks. Croft et al.~\cite{croft2023data} investigated the significance of data quality in software vulnerability datasets and revealed that inaccuracies in vulnerability labels and duplicated data points can potentially lead to ineffective model training and unreliable results. Wu et al.~\cite{wu2021data} discussed the adverse effects of inaccurate dataset labels on prediction decisions, and Chakraborty et al.~\cite{chakraborty_deep_2020} identified that duplicate data, inadequate token-based models, irrelevant feature learning, and imbalanced data cause poor performance of existing models when applied to real-world problems.
A study conducted by Herbold et al.~\cite{herbold_fine-grained_2021} that manually validates a large corpus of bug fixing commits, highlighted the limitations of manual validation and the impact of bug-irrelevant lines on data quality.
The granularity of the existing vulnerability related datasets varies from file-level~\cite{yu2019improving} to function level~\cite{chakraborty_deep_2020, russell_automated_2018, zheng_d2a_2021, fan_cc_2020}.
Reveal~\cite{chakraborty_deep_2020} uses publicly available patch. Devign~\cite{zhou_devign_2019} uses manually labeled data from commits filtered through keywords, and Big-Vul~\cite{fan_cc_2020} employs the description of Common Vulnerabilities and Exposures (CVE) database as well as commits to generate function-level vulnerability dataset. Yu et al.~\cite{yu2019improving} suggested using active learning with user-feedback to improve file-level vulnerability predictions. Compared to our work, these works lack the granularity required to train a good line-level vulnerability detection tool. At line-level granularity, BugBuilder~\cite{jiang_bugbuilder_2023} and D2A~\cite{zheng_d2a_2021} use test cases and differential static analysis to exclude irrelevant changes. However, both these approaches are limited in types of vulnerability they can process due to their filtering methods. In comparison, our approach learns to distinguish between bug-relevant and bug-irrelevant lines using patterns in the extracted features and would be free from such restrictions.
In the past, there have been many works on processing tangled commits. These works have used testing~\cite{yang_is_2021, hashimoto_automated_2018}, graph clustering~\cite{li2022utango,chen_untangling_2022,shen_smartcommit_2021}, or feedback based tools~\cite{wang_cora_2019,kirinuki2014hey}  to identify tangled changes. DEPTEST~\cite{yang_is_2021} and Delta Debugging~\cite{hashimoto_automated_2018} used automated testing to filtering tangled code changes. Kirinuki \cite{kirinuki2014hey} and CoRA~\cite{wang_cora_2019} created tools that warned or helped users avoid making tangled changes in the first place.
ComUnt~\cite{chen_untangling_2022}, SmartCommit~\cite{shen_smartcommit_2021}, UTango~\cite{li2022utango}, and Flexeme~\cite{partachi_flexeme_2020} address this issue with varied graph-based approaches, including graph partitioning. %

%% file: conclusion.tex
\section{Conclusion And Future Work}\label{sec:CleanConclusion}

This paper presents \tool{}, an automatic and scalable tool for generating line-level vulnerability data. \tool{} uses active learning to significantly reduce the labeled data needed to train the model. We designed features considering commit lines and their surrounding code. In our evaluation, we used both Java and C datasets and processed more than 4.3K commits and 119K commit lines. We show that commits are very noisy and many commits can contain such noise. \tool{} is able to find such lines that are not relevant to vulnerability, and reported F1 scores between 70--74. Using \tool{}, we generated line-level vulnerability data for \texttt{FFmpeg} in the Devign dataset.  We demonstrated that using this cleaned dataset, LineVul is able to detect 18 more vulnerable functions at 87 F1 score and 70 more vulnerable lines at 73\% top 10, compared to 83 F1 scores and 66\% top 10 reported by the baseline. By sampling 50 labels recommended by \tool{} for correction, we find 29 incorrect function-level labels for \texttt{FFmpeg}.  In the future, we will use \tool{} to continue generating more line-level datasets and reporting false labels in the current vulnerability datasets.
